\documentstyle[11pt,aaspp4]{article}

\newcommand{\etal}{{\it et al.\/}}

\begin{document}

\title{The Extremely Red Objects Found Thus Far in the
Caltech Faint Galaxy Redshift Survey\altaffilmark{1}}

\author{Judith G. Cohen\altaffilmark{2},
        David W. Hogg\altaffilmark{3,4,5}
        Roger Blandford\altaffilmark{3},
        Michael A. Pahre\altaffilmark{2,5,6} \&
        Patrick L. Shopbell\altaffilmark{2} }

\altaffiltext{1}{Based in large part on observations obtained at the
	W.M. Keck Observatory, which is operated jointly by the California 
	Institute of Technology and the University of California}
\altaffiltext{2}{Palomar Observatory, Mail Stop 105-24,
	California Institute of Technology, Pasadena, CA \, 91125}
\altaffiltext{3}{Theoretical Astrophysics, California Institute of Technology,
	Mail Stop 130-33, Pasadena, CA \, 91125}
\altaffiltext{4}{Current Address:  Institute for Advanced Study, Olden Lane, Princeton, NJ \, 08540}
\altaffiltext{5}{Hubble Fellow}
\altaffiltext{6}{Current Address:  Harvard-Smithsonian Center for Astrophysics, 
	60 Garden St., Mail Stop 20, Cambridge, MA \, 02138}

\begin{abstract}
We discuss the very red objects found in the first field
of the Caltech Faint Galaxy Redshift Survey, for which the
observations and analysis are now complete.  In this field,
which is 15 arcmin$^2$ and at J005325+1234 there are 195 objects
with $K_s < 20$ mag, of which 84\% have redshifts.
The sample includes 24 spectroscopically confirmed Galactic stars,
136 galaxies, three AGNs, and 32 objects without redshifts.

About 10\% of the sample has $(R-K) \ge 5$ mag.  Four of these objects
have redshifts, with $0.78 \le z \le 1.23$.  Three of these are 
based on absorption features in the mid-UV, while the lowest
redshift object shows the standard features near 4000\AA.   Many of the
objects still without redshifts have been observed spectroscopically, 
and no emission lines were seen in their spectra.  We believe they 
are galaxies with $z \sim 1 - 1.5$ that are red due to their age 
and stellar content and not to some large amount of internal 
reddening from dust.

Among the many other results from this survey of interest here is
a determination of the median extinction in the mid-UV for
objects with strong emission line spectra at $z \sim 1 - 1.3$.
The result is extinction by a factor of $\sim$2 at 2400\AA.

\end{abstract}

\section{Introduction}

We have completed the analysis of the data for the first field
of this survey, which is 2 x 7.3~arcmin$^2$ field at J005325+1234.
The sample is selected ignoring morphology
at $K$ and consists of the 195 objects with
$K < 20$ mag in this field.  These were observed with the Low
Resolution Imaging Spectrograph (Oke \etal\ 1995) 
at the Keck Observatory.  Six color
photometry ($UBVRIK$) is available for the entire field as well
from Pahre \etal\ (1998).

Redshifts were successfully obtained for 163 of the 195 objects in the 
sample to achieve a completeness of 84\%.
These redshifts
lie in the range [0.173, 1.44] and have a median of 0.58 
(excluding 24 spectroscopically confirmed Galactic stars).
The sample includes two broad lined AGNs and one QSO.
The objects are assigned to spectral classes based on the
relative preponderance of emission lines versus absorption lines
in their spectra.  The four spectral classes used for extragalactic
objects are ``${\cal E}$'' for emission line dominated spectra (33 galaxies),
``${\cal A}$'' for absorption line dominated spectra (51 galaxies), 
``${\cal C}$'' for composite spectra (52 galaxies), and ``${\cal Q}$''
for AGNs.  A few starbursts were found, classified as ``${\cal B}$'',
but for the present discussion they are grouped together with the
emission line galaxies.

\section{Rest Frame Spectral Energy Distributions}

\begin{figure}
\plotone{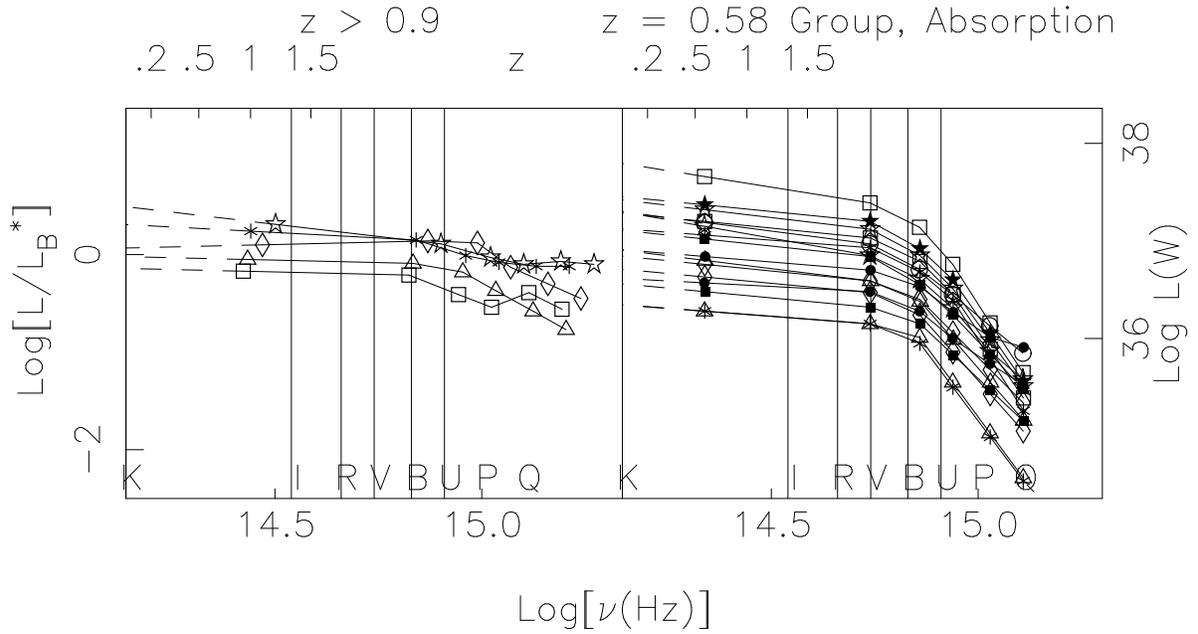}
\caption{ The rest frame spectral energy
distributions (SEDs) for selected galaxies.
The abscissa is the rest frequency and 
the rest wavelengths corresponding to our 6 color 
photometry augmented by the two supplementary ultraviolet bands 
$P$ and $Q$ (log($\nu$)=15.0 and 15.1) are indicated.  The ordinate
is the logarithm of the spectral power in units of both 
$L_B^\ast$ and W.  Each galaxy
SED shows the rest wavelengths corresponding to the observations 
and dashed lines are used
to indicate extrapolations.  The upper horizontal scale can be used in conjunction with the $K$ point
to measure the redshift of the galaxy.  }
\label{fig-1}
\end{figure}

The galaxy rest frame SEDs derived from out $UBVRIK$ photometry 
are very closely
correlated to the galaxy spectral types.  Both are also correlated
with galaxy luminosity; blue galaxies show the signature of recent
star formation in their spectra and are less luminous for $z < 0.8$ than
red galaxies which show no evidence for recent star formation in
their spectra.  Representative SEDs are shown in Figure~\ref{fig-1}.
The SEDs for selected galaxies 
(D0K183, 172, 108, 188 and 158) 
with $z > 0.9$ shown in Figure~1a are remarkably flat (blue). 
Figure~1b shows the SEDs for all the absorption
line galaxies in the $z = 0.58$ peak; they have quite steep (red) spectra.

\subsection{The Extremely Red Objects in Our Sample}

\begin{figure}
\plotone{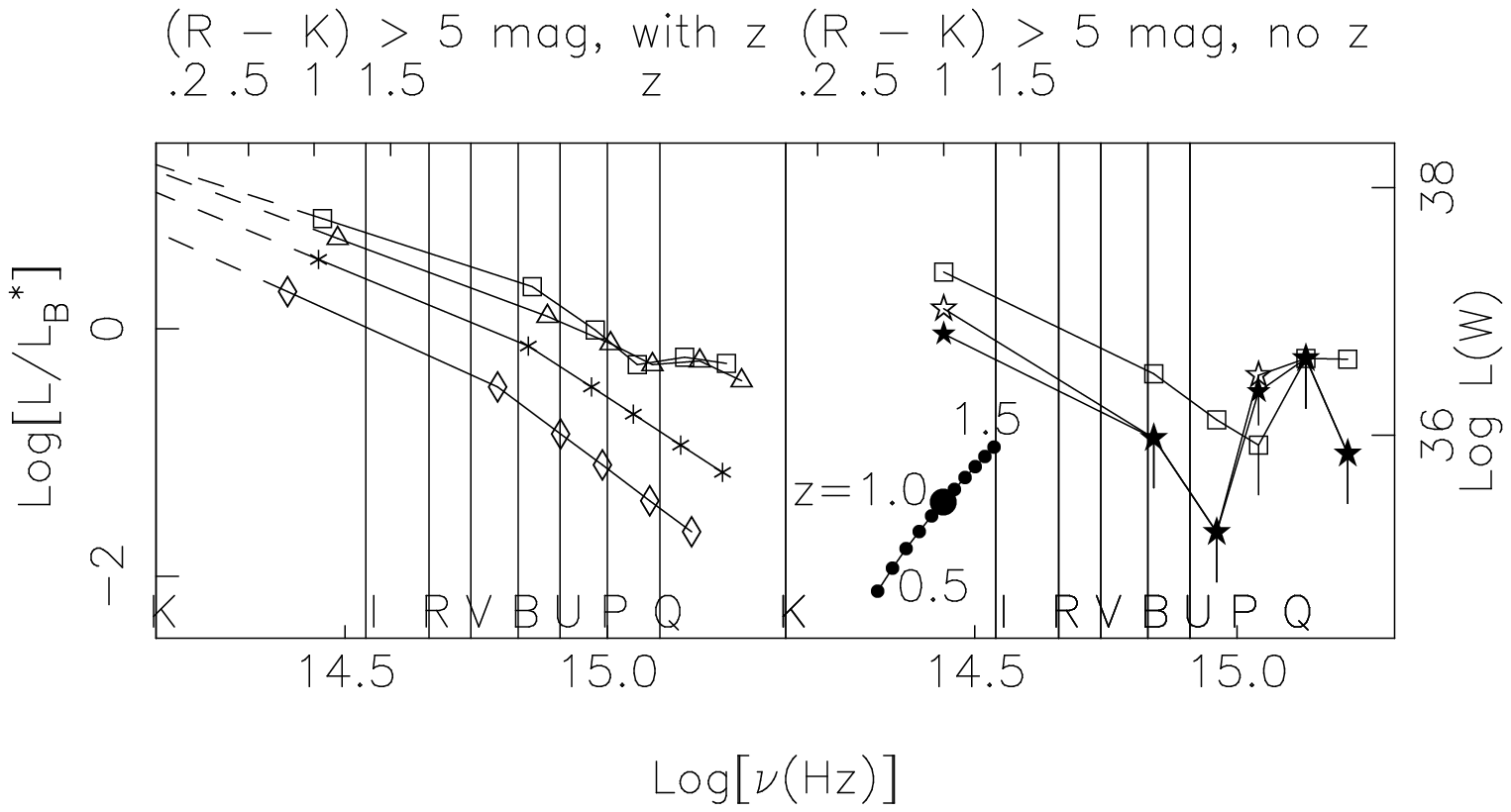}
\caption{The rest frame SEDs for the four extremely red galaxies for which
redshifts have been determined from our survey.  The second panel shows the
rest frame SEDs for three of the extremely red galaxies without
redshifts, calculated assuming $z = 1$. 
The line in the lower left indicates how the SEDs will shift for
$0.5<z<1.5$. }
\label{fig-2}
\end{figure}

There are 24 Galactic stars in this sample, mostly M dwarfs or
M subdwarfs.  The reddest galactic star identified
spectroscopically in this field has $(R-K) = 4.6$ mag.
There are 19 objects in this sample with $(R-K) \ge 5$ mag, which
we call the very red objects, and which we believe to be galaxies rather
than Galactic stars. Four of these have redshifts, most of
which are somewhat uncertain.
Figure~\ref{fig-2} shows the rest frame SEDs for the four very 
red galaxies with redshifts.

The second panel of Figure~\ref{fig-2} shows the SEDs for three of the 
very red objects which do not have redshifts, calculated assuming
$z = 1$.  Redder than $B$, these look similar
to those SEDs shown in the first panel of this figure, 
but the objects are somewhat fainter.
Most of the $U$ and $B$ magnitudes
for these objects are upper limits, as indicated by the vertical bars
going downward from the relevant points.

A more complete discussion of the redshift peaks (i.e. groups
and poor clusters of galaxies), luminosity function, 
the cosmological volume density, the constraints on mergers,
the ultraviolet extinction and other
issues can be found in two papers, one of which
has been submitted to \apj\ (Cohen \etal 1998a) while the other
(Cohen \etal\ 1998b) will be published in \apjs.

\section{Final Comments}

We have determined the fraction of very red objects among our sample.
For counts to $K < 20$ mag, $\sim$10\% of the sample of 195 objects
is very red, i.e. has $(R-K) \ge 5$ mag.  If one excludes the 
known Galactic stars from the sample, this fraction does not 
change substantially.

We have examined the spectra of many of these extremely red objects
and have succeeded in determining the redshifts of four of them,
although the redshifts are not as certain as one might desire.
We suggest that these are galaxies with $z \sim 1 - 1.5$ 
in which reddening by dust is not playing a major role.
In particular they are not heavily reddened starbursts.
(If they were, we should have
seen some moderately reddened emission line galaxies, and there
were no such beasts among our sample).  Instead we believe
their extremely red colors are a direct consequence of their age, 
stellar composition , k-corrections, etc. and that these
extremely red objects 
are the analogs at this redshift range of local elliptical galaxies.
We thus support Persson \etal\ (1993)  and Graham \& Dey (1996), who
among others, 
have speculated that such objects are passively-evolved elliptical 
galaxies with $z > 1$.

More work is going to be required to get some first class redshifts
for these, or similar, hopefully brighter, objects, to establish
their nature in a more definitive way.

\acknowledgements

The entire Keck/LRIS user community owes a huge debt
to Jerry Nelson, Gerry Smith, Bev Oke, and many other people who have
worked to make the Keck Telescope and LRIS a reality.  We are grateful
to the W. M. Keck Foundation, and particularly its late president,
Howard Keck, for the vision to fund the construction of the W. M. Keck
Observatory. JGC is grateful for partial support from STScI/NASA grant AR-06337.12-94A.
DWH and MAP were supported in part by Hubble Fellowship grants 
HF-01093.01-97A and HF-01099.01-97A from STScI (which is operated 
by AURA under NASA contract NAS5-26555).

\end{document}